\documentclass[aps,prper,twocolumn,showpacs,letterpaper,longbibliography]{revtex4-1}

\usepackage{graphicx}
\usepackage{mdframed}
\usepackage{framed}
\usepackage{indentfirst}
\usepackage{natbib}
\usepackage{amsmath,amssymb}
\usepackage{subfigure}
\usepackage{enumitem}
\usepackage{epstopdf}
\usepackage{xfrac}
\usepackage{multirow}

\usepackage{changes}

\begin{document}

\title{A cross-context look at upper-division student difficulties with integration}

\pacs{01.40.Fk}
\keywords{physics education research, upper-division, laboratory, attitudes, assessment, instruction}

\author{Bethany R. Wilcox}
\affiliation{Department of Physics, University of Colorado Boulder, Boulder, CO 80309}

\author{Giaco Corsiglia}
\affiliation{Department of Physics, University of Colorado Boulder, Boulder, CO 80309}

\begin{abstract}
We investigate upper-division student difficulties with direct integration in multiple contexts involving the calculation of a potential from a continuous distribution (e.g., mass, charge, or current).  Integration is a tool that has been historically studied at several different points in the curriculum including introductory and upper-division levels.  We build off of these prior studies and contribute additional data around student difficulties with multi-variable integration at two new points in the curriculum: middle-division classical mechanics, and upper-division magnetostatics.  To facilitate comparisons across prior studies as well as the current work, we utilize an analytical framework that focuses on how students activate, construct, execute, and reflect on mathematical tools during physics problem solving (i.e., the ACER framework).  Using a mixed-methods approach involving coded exam solutions and student problem-solving interviews, we identify and compare students' difficulties in these two different context and relate them to what has been found previously in other levels and contexts.  We find that some of the observed student difficulties were persistent accross all three contexts (e.g., identifying integration as the appropriate tool, and expressing the difference vector), while other difficulties seemed to fade as students advanced through the curriculum (e.g., expressing differential line, area, and volume elements).  We also identified new difficulties that appear in different contexts (e.g., interpreting and expressing the current density). 
\end{abstract}

\maketitle

\section{\label{sec:intro}Introduction}

Physics Education Research (PER) as a field has a long history of conducting student difficulties research focused on identifying the challenges and problems that students face when dealing with specific physics concepts or mathematical tools \cite{hsu2004problemsolving, meltzer2012resource}.  The findings from these studies have historically been used to inform the development of curricular materials, pedagogical approaches, or classroom practice in order to specifically target these challenging areas to directly address and overcome known student difficulties \cite{meltzer2012resource, chasteen2015sei}.  One clear example of this process was the creation of the ``Tutorials in Introductory Physics" \cite{mcdermott2001tutorials}, which were developed based on extensive investigations of specific student difficulties with introductory physics content and have been show to improve student learning gains \cite{finkelstein2005tutorial}.

Historically, studies of student difficulties have been performed in relation to a single concept, topical area, or mathematical tool and within the context of a single course.  However, the cyclic nature of the physics curriculum means that students often encounter individual concepts and mathematical tools multiple times, and in multiple contexts, over the course of their undergraduate career.  It is reasonable to assume that over this time period, and with multiple exposures, students’ difficulties with these concepts and tools may naturally evolve.  For example, some difficulties faced early on may resolve themselves over time as students see these ideas multiple times in the standard curriculum and gain more experience, while other difficulties may be more persistent, reappearing despite repeated exposures.  Alternatively, as the complexity of the physics content associated with these topics and tools increases, new difficulties not present in earlier contexts may also begin to appear.

Understanding the nature of how students’ difficulties with common concepts and mathematical tools change as students are exposed to these ideas in new contexts can have important implications for both instructors and researchers.  By attending to how students' ideas evolve, we can avoid spending time and effort on developing new materials to target difficulties that are already being sufficiently addressed in the standard curriculum in favor of focusing on difficulties that persist across multiple contexts and exposures.  Yet, despite the potential advantages, investigations of student difficulties with specific topics or tools across multiple exposures is rare within the PER literature.  In lieu of comprehensive investigations looking at the evolution of student difficulties over time, another option for investigating this dynamic would be to make explicit comparisons between distinct studies investigating the same topic or tool at different levels of the curriculum.  However, such cross-study comparisons would be difficult without a guiding framework to facilitate making connections across different studies.  One framework that was created to provide just such a guiding structure is the ACER framework \cite{caballero2015mathphys}, which focuses on how students Activate, Construct, Execute, and Reflect on mathematical tools in the context of physics problem solving \cite{wilcox2013acer}.

One example of student difficulties research that focuses on a mathematical tool that appears multiple times throughout the undergraduate curriculum is direct integration.  Here, use ``direct integration" to refer to the process of calculating a physical quantity (e.g., the gravitational potential) by directly summing the contributions to that quantity from each `bit' of a physical distribution (e.g., a three dimensional mass distribution).  We build on the existing research on students difficulties with direct integration to construct a pseudo-longitudinal and cross-context view of how students' difficulties evolve over time.  To accomplish this, we first review the existing literature on students’ difficulties with direct integration at all levels of the curriculum (Sec.\ \ref{sec:lit}).  We then build on this work by presenting methods (Sec.\ \ref{sec:methods}) and findings (Sec.\ \ref{sec:results}) from a pseudo-longitudinal investigation of student difficulties with direct integration in middle-division classical mechanics (in the context of gravitation) and upper-division electricity and magnetism (in the context of both the scalar and vector potential).  Throughout this paper, the ACER framework provides a consistent structure that facilitates making comparisons across topical areas as well as to previous studies.

\section{\label{sec:lit}Background: Student difficulties with direct integration}

Here, we discuss prior work investigating student difficulties with direct integration.  Throughout this section, the ACER framework will be used facilitate comparisons of findings across these diverse studies.  As such, we begin with a summary of our own prior work using the ACER framework to investigate students use of direct integration in the context of upper-division electrostatics and then continue to discuss comparisons with work in other contexts and at other levels.

\subsection{\label{sec:acer}The ACER framework and an example in upper-division electrostatics}

The ACER framework is an analytical framework developed to help structure and organize investigations of students' difficulties when utilizing sophisticated mathematical tools in the context of physics problem solving \cite{wilcox2013acer}.  The framework considers four main components present in this type of mathematically involved problem solving: \emph{activation} of the mathematical tool, \emph{construction} of the mathematical model, \emph{execution} of the mathematics, and \emph{reflection} on the results.  These components were identified by studying the general structure of expert problem solving using modified task analysis\cite{catrambone2011taps}.  The framework itself is grounded in both a resources \cite{hammer2000resources} and an epistemic framing \cite{bing2008thesis} perspective on the nature of knowledge and the process of learning.  While the broad structure of the ACER framework was designed to be applicable across contexts and mathematical tools, the framework was also designed to be operationalized for use with specific topics and tools.  This operationalization process is important to ensure that the framework captures the more tool- and context-specific aspects of students' difficulties.  The specific operationalization of the ACER framework for the contexts of gravitation and the vector potential will be discussed in greater detail in Sec.\ \ref{sec:oper}.  For the purposes of cross-study comparisons, we will focus categorizing students' difficulties within the general components of the framework (i.e., activation, construction, execution, and reflection).

One of the investigations which prompted and informed the development of the ACER framework was that of direct integration in the context of upper-division electrostatics, specifically the calculation of the electric potential via the integral form of Coulomb's Law \cite{wilcox2013acer}. This initial study laid the foundation for the work described in the remainder of this manuscript and features very similar contexts and methodologies including quantitative analysis of students' exam solutions as well as qualitative analysis of interview data.  From this analysis, we identified two broad categories of difficulties that the study population exhibited with integration in the context of Coulomb's law.  The first was difficulty identifying direct integration as the appropriate solution method (i.e., difficulty with the \emph{activation} component).  This difficulty was exhibited by roughly a quarter of the student population \cite{wilcox2013acer}.  The second was difficulty operationalizing the Coulomb's law integral for the specific physical situation given (i.e., difficulty with the \emph{construction} component).  The two most common difficulties observed were with correctly expressing the differential charge element, $dq$, and the difference vector (i.e., Griffith's ``script-r") in a way consistent with the geometry of the given physical situation.  These two difficulties were each exhibited by roughly half of the students \cite{wilcox2013acer}.  Additionally, we found that very few students in our population showed explicit and spontaneous attempts to interpret or check their solutions (i.e., operating within the \emph{reflection} component) \cite{wilcox2013acer}.

\subsection{\label{sec:otherLit}Interpreting other prior work through the ACER framework}

In addition to our own prior work investigating student difficulties with direct integration in the context of upper-division electrostatics, a number of others have investigated student difficulties in introductory courses.  Here, we summarize this prior work from the perspective of the ACER framework to facilitate cross-study comparisons.  It is important to note that these studies were not designed or conducted using the ACER framework, and often had goals that went beyond pure investigations of students' difficulties (e.g., theoretical development such as understanding students' resource activation or conceptual blending); however, for the purposes of this summary, we focus on the aspects of these findings that directly relate to student difficulties with integration as a mathematical tool in physics problem solving.

Many of the existing studies on student difficulties with integration were conducted in the context of introductory physics courses.  For example Nguyen and Rebello \cite{nguyen2011int} examined student difficulties when calculating the electric field from a straight or curved line segment.  They focused explicitly on understanding how students recognized the need for integration and how they set up and then computed the integrals \cite{nguyen2011int}.  From the point of view of ACER, this framing aligns well with a focus on the \emph{activation} and \emph{construction} components of the framework as well as, potentially, the \emph{execution} component.  With respect to \emph{activation}, Nguyen and Rebello found that students often used simple recall of similar problems to activate integration as the correct mathematical tool.  In questions where they had less experience with similar problems, students had more difficulty identifying the need for a integral, and those who did were most often cued by the presence of a non-constant value in the prompt.  With respect to \emph{construction}, the two primary difficulties focused around the infinitesimal quantity and accounting for the vector nature of the electric field.  Nguyen and Rebello noted that many students struggled to recognize the need for, and physical meaning of, the infinitesimal quantity as well as to express it in a useful way (e.g., expressing $dq$ as $\lambda d x $).  Finally, with respect to \emph{execution}, Nguyen and Rebello documented some difficulties related to maintaining an awareness of the physical meaning of the symbols while performing the integrals.  They also noted some difficulties with respect to the actual mechanics of computing the integrals (e.g., u-substitutions).

Meredith and Marrongelle \cite{meredith2008resources} also conducted an interviews with students solving several problems requiring integration in the context of introductory electrostatics, including one asking for the electric field near a bar of charge.  The focus of this work was on identifying students' mathematical resources when solving integrals; however, they also documented a number of difficulties that their interview students had working through these types of questions.  In terms of \emph{activation}, they found that recognizing the need for an integral in the case of the bar of charge was a challenge for as many as half of their students, with many of these student instead treating the bar as a point charge at the rod's center.  For students who successfully identified the need for integration, Meredith and Marrongelle identified three possible cues commonly used: recall of prior examples, identification of a term or variable on which the quantity in question depended, or the conceptualization of a building up a whole by summing up smaller constituent parts.  In terms of \emph{construction},  Meredith and Marrongelle also observed difficulties with the infinitesimal, finding that students sometimes incorrectly used the idea of dependence on a particular variable to simply assume that would be the variable of integration without consideration of the physical meaning of the integrand.

Both of the two studies discussed above identified difficulties with the infinitesimal/differential term in the larger context of solving a physics problem involving integration, an element that appears in the \emph{construction} component of the ACER framework.  Others have targeted this issue more directly.  For example, Hu and Rebello \cite{hu2013differentials} conducted an interview study with introductory physics students in which they focused specifically on students applications of differentials when calculating the electric field from a charged bar.  They identified different mathematical resources and conceptual metaphors that students used when considering the need for, and physical meaning of, the differential term in these integrals.  Similarly, Amos and Heckler \cite{amos2018differentials} investigated the relationship between introductory students' understanding of differentials (e.g., $dx$), differential products (e.g., $\lambda dx$), and integrals.  They found that understanding of differentials alone did not relate strongly to performance on integral problems, but rather required an additional step of explicitly incorporating ideas about differential products.

Others, rather than focusing on the \emph{construction} component, focused instead on the \emph{activation} component through investigations of how students identify integration as the correct mathematical tool for a particular physics problem.  To investigate this, Savelsbergh \emph{et.\ al} \cite{savelsbergh2011approach} used problem sorting tasks with introductory physics students which asked them to determine the correct approach to various problems relating to electricity and magnetism; several of these tasks included problems for which direct integration via Coulomb's law or the Biot-Savart Law were the correct approach.  They found that, despite knowing the possible solution types for these problems, students had difficulty mapping the problem at hand onto one of the know problem types.  In terms of the ACER framework, these results would indicate that even if a student can productively recognize or replicate work in the \emph{construction} or \emph{execution} components, they may still struggle to identify the correct mathematical tool in the \emph{activation} component.

Taken together, the studies described above tell a relatively coherent story with respect to the difficulties students at the introductory level face with multi-variable integration in the context of, for example, calculating the electric field from a charge distribution.  These difficulties are focused on the \emph{activation} and \emph{construction} components of the ACER framework and emphasize difficulties identifying integration as necessary and manipulating the differential term.  Both of these difficulties also appear in our own prior work investigating student difficulties with these same types of calculations in the context of upper-division electrostatics, suggesting that these difficulties are relatively persistent across several exposures to the use of multi-variable integration in physics problems.  However, as the majority of the studies at the introductory level are based primarily on student interview data, conclusions about potential changes in the frequency of these difficulties between the introductory and upper-division population are not possible.  Missing from the majority of these prior studies is explicit attention to students' \emph{reflections}, which is also consistent with our finding that students rarely exhibit reflective behaviors spontaneously.

The current study builds on this prior work by contributing data on students difficulties at an additional two points within the undergraduate curriculum: middle-division classical mechanics (in the context of gravitation) and upper-division magnetostatics (in the context of the vector potential).  Additionally, we will leverage both quantitative exam data and qualitative interview data, along with comparison to our prior work, to make statements about how the frequency of these difficulties shifts (or not) over the course of these multiple exposures.

\section{\label{sec:methods}Methods}

\subsection{\label{sec:data}Context}

Data for this study were collected from two courses at the University of Colorado Boulder (CU) over the course of three distinct semesters.  Both courses in this study are the first in a two semester sequence - one targeting a combination of classical mechanics and math methods (Class.\ Mech.), and the other targeting electricity and magnetism (E\&M).  Class.\ Mech., covers chapters 1-5 in Taylor's text \cite{taylor2005classmech} as well as various chapters in Boas \cite{boas2006math}, and students are typical sophomores and juniors.  E\&M, covers the first 6 chapters of Griffths' text \cite{griffiths1999em} (i.e., electrostatics and magnetostatics), and students in this course are juniors and seniors.  For both courses, most students are physics, astrophysics, and engineering physics majors. During the semesters of data collection, both courses were taught with varying degrees of interactivity through the use of research-based teaching practices including peer instruction using clickers \cite{mazur1997pi} and tutorials \cite{chasteen2012transforming}.

We collected data from two primary sources for this investigation: student solutions to instructor-designed questions on traditional midterm exams, and group think-aloud, problem-solving interviews.  Our approached mirrors that of our earlier investigations of students' difficulties with direct integration described in the prior section (Sec.\ \ref{sec:acer}).  In this approach, exams provided quantitative data identifying common difficulties and interviews offered deeper insight into the nature of those difficulties.  At CU, Class.\ Mech.\ is a pre-requisite for E\&M, and, thus, students' responses in these two courses during the same semester represent a pseudo-longitudinal view of the evolution of students' reasoning over the course of several semesters.  These data are pseudo-longitudinal, rather than truly longitudinal, because rather than tracking individual students and comparing their performance at different points in time, we are comparing the aggregate performance of two different sets of students at different points in the curriculum.  The pseudo-longitudinal nature of our data has implications for the interpretation of this data that will be discussed in more detail in the following section (Sec.\ \ref{sec:results}).

\begin{figure*}
 \subfigure[]{
        \begin{minipage}{0.45\linewidth}
        \begin{mdframed}\vspace*{2mm}
                	\center 
					\includegraphics[width=0.60\linewidth]{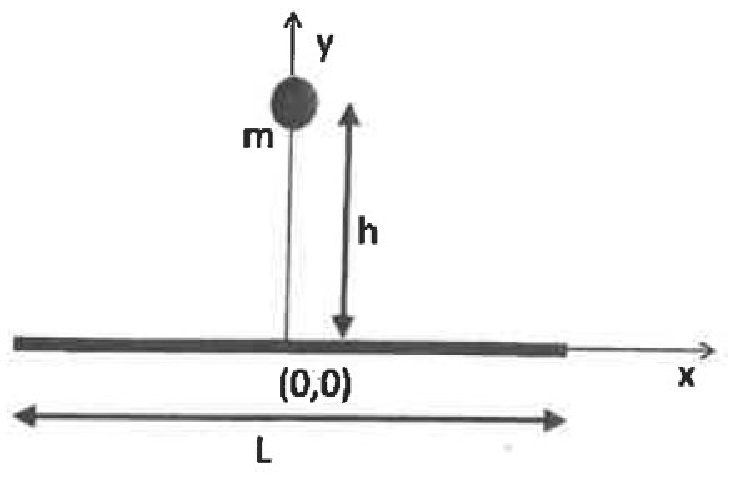}
				\flushleft A thin rod of length $L$ lies along the x-axis, with its center at the origin, as shown in the figure.  The linear mass density of the rod $\lambda$ is a constant.  A mass $m$ is a vertical distance $h$ above the center of the rod. \\ \vspace*{2mm}
					Calculate the gravitational potential energy of the mass $m$ due to the mass of the rod.  \\
				\vspace*{18mm}
        \end{mdframed}
        \end{minipage}
        \label{fig:cm1}}      
        \subfigure[]{
		\begin{minipage}{0.45\linewidth}
        \begin{mdframed}\vspace*{2mm}
               	\center 
					\includegraphics[width=0.450\linewidth]{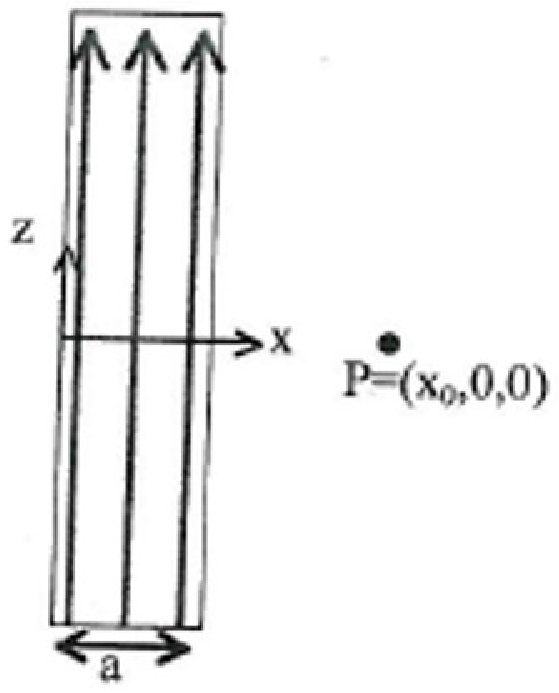}
				\flushleft Consider a flat ``ribbon'' of current of width $a$ and length $b$, flowing in the $x$-$z$ plane as shown.  This is a surface current in the $y=0$ plane with\\
					 \[ {\bf K}= \begin{cases}
										K_o \hat{z} & where \hspace{1mm} 0<x<a \\
										0  & where \hspace{1mm} x<0 \hspace{1mm} or \hspace{1mm} x>a \\ 
								\end{cases} \] 
				\flushleft Find an expression for the magnetic vector potential, $A(x_o,0,0)$, at the point P shown.   \\

 				\vspace*{2mm}
        \end{mdframed}
        \end{minipage}
        \label{fig:em1}}
	\caption{Examples of the exam questions used in the study.  Variations on (a) included asking instead for the scalar potential from a short rod of charge or for the vector potential from a short line segment of current.  } \label{fig:equestions}
\end{figure*}

Midterm exam data were collected from one semester in Class.\ Mech.\ ($N=77$ students and 1 instructor) and two semesters in E\&M ($N=163$ distinct students and 3 instructors).  Of the four total instructors involved in these courses, two were traditional research faculty and two were physics education researchers. One semester of the E\&M course was co-taught by one PER faculty member and one of traditional research faculty with shared exams, activities, and homework.  The exam questions used for the study were developed via collaboration between the traditional research faculty and the PER faculty.  The exam questions were all variations of calculating the potential from a short line or strip segment (see Fig.\ \ref{fig:equestions}).  For example, in Class.\ Mech.\, the students were asked for the gravitational potential energy from a rod of linear mass density.  The E\&M students were asked two questions: one asking for the scalar potential from a short charged rod, and one asking for the vector potential from a short line carrying constant current current or a short strip carrying uniform surface current density.  The most efficient solution approach in each case is to calculate the requested potential directly using Eqns.\ \ref{eqn:u}, \ref{eqn:v}, or \ref{eqn:a}.  

\begin{equation}
U(\vec{r}) = -GM\int_V \frac{dm}{|\vec{r}-\vec{r}'|} \label{eqn:u}
\end{equation}
\begin{equation}
V(\vec{r}) = \frac{1}{4 \pi \epsilon_0}\int_V \frac{dq}{|\vec{r}-\vec{r}'|} \label{eqn:v}
\end{equation}
\begin{equation}
\vec{A}(\vec{r}) = \frac{\mu_o}{4 \pi}\int_V \frac{\vec{J}}{|\vec{r}-\vec{r}'|}dV' \label{eqn:a}
\end{equation}

\noindent Here $G$, $\epsilon_o$, and $\mu_o$ are fundamental constants, $M$ is the mass of the of the test object, $\vec{r}$ is the vector from the origin to the field point, and $\vec{r}'$ is the vector from the origin to the source point.

While all exam questions in the study were some variation on calculating the potential from a short rod or strip, there were several important differences between the prompts used in each case.  All questions provided a set of Cartesian axes in the figure; however, the orientation of the distribution (i.e., what axis it lay along) varied between prompts.  More significantly, one version of the two vector potential questions provided students with the correct mathematical tool to use (i.e., Eqn.\ \ref{eqn:a}), whereas in all other exam questions students had to decide for themselves which tool to use.  From the point of view of ACER, this means that this specific vector potential prompt bypassed the \emph{activation} component. Additionally, only the two scalar potential prompts resulted in integral expressions that could actually be calculated by hand; in the other cases, students were either asked only to set up the integral or were provided the result of the integral.  With respect to ACER, this means that the questions targeting gravitation and the vector potential all effectively bypassed the \emph{execution} component.  

\begin{figure}[b]
\begin{mdframed}
	\flushleft It can sometimes be useful to model electrons in orbitals around an atom as small rings of current.  In the figure below, we have provided a diagram of a ring carrying current I in the counter-clockwise direction as viewed from above.  Calculate the vector potential along the x-axis from this ring of current.  \\
	\center \includegraphics[width=0.6\linewidth]{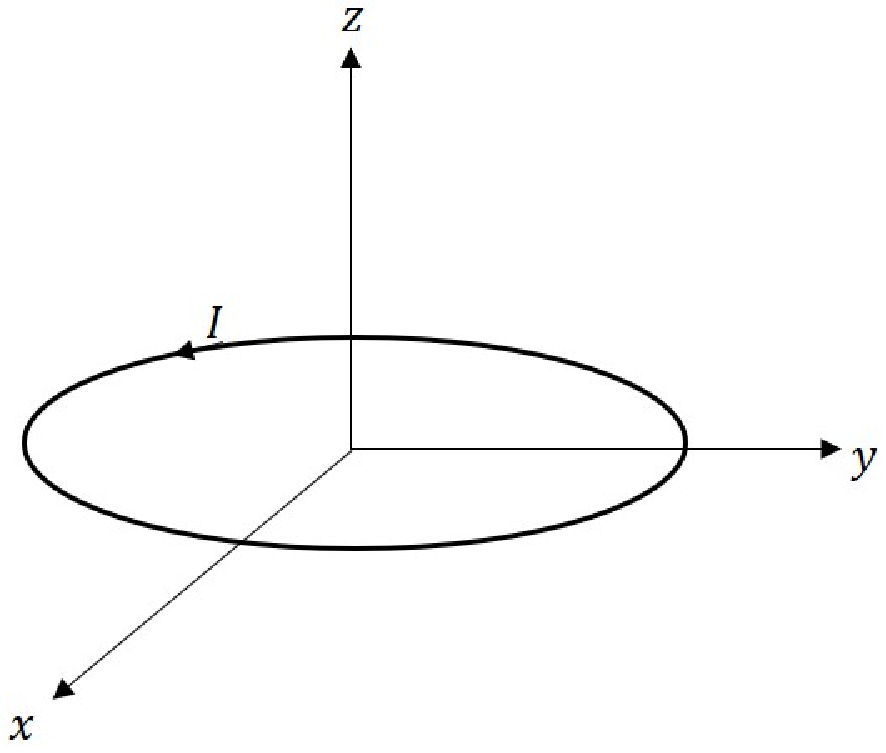}
\end{mdframed}
\caption{The vector potential version of the interview question.  For the Class.\ Mech.\ students, the distribution was described as a mass distribution with uniform $\lambda$ (and the arrow indicating direction of current was removed), and students were asked to calculate the gravitational potential energy for a small mass placed on the $x$-axis.  } \label{fig:iquestion}
\end{figure}

To address the lack of explicit \emph{reflection} and other limitations with the exam data, we also conducted think-aloud interviews with pairs of students.  Four interviews were conducted with students ($N=8$) who had previously completed or were currently enrolled in the Class.\ Mech.\ course, and three interviews were conducted with students ($N=6$) who had previously completed, or were currently enrolled in, the E\&M 1 course.  In these interviews, students were asked to respond to a slightly more challenging version of the exam questions, which asked them to calculate the potential along an axis other than the central axis from a ring of uniform density (see Fig.\ \ref{fig:iquestion}).  The increase in difficulty in the interviews was to ensure the question was sufficiently challenging for a pair of students working together.  The paired interview structure was adopted to create a more authentic interview environment in which interview participants would naturally encourage each other to express their reasoning without prompting from the interviewer.  In both interview sets, students were not told which mathematical tool to use; however, a correct solution to the problem will produce an integral that is not easily evaluated by hand.  As such, interviewees who reached this integral were asked to stop at that point.  From the perspective of the ACER framework, these interview questions targeted \emph{activation}, \emph{construction}, and potentially \emph{reflection}, but did not capture the majority of the \emph{execution} component.  This design was based on findings from our prior work suggesting that pure \emph{execution} errors are rarely the primary difficulties encountered by students \cite{wilcox2013acer}.

The exams solutions contained little evidence of spontaneous \emph{reflection} (see Sec.~\ref{sec:reflection}).  In one semester, the exam question asking for the scalar potential include two follow up questions asking students to discuss the limiting behavior of the potential in the large-$r$ limit and to confirm that their solution exhibited this behavior.  The interviews also included an additional question designed to examine students' ability to come up with meaningful reflections when prompted to do so.  Five of the seven pairs (those with interview time remaining after finishing the first question) completed this second question targeted directly at \emph{reflection}.  Students were asked to assess the plausibility of three expressions for either gravitational potential energy or vector potential caused by an unspecified, but localized, mass or current distribution.  Valid approaches included (but are not limited to) checking units or limits, but neither approach was suggested explicitly.  

\subsection{\label{sec:oper}The operationalized framework}

The process of operationalizing ACER is presented in detail in Ref.\ \cite{wilcox2013acer}.  Briefly, in order to operationalize the framework, a content expert utilizes a modified form of task analysis \cite{catrambone2011taps, wilcox2013acer} in which they work through the problems of interest while carefully documenting their steps and mapping these steps onto the general components of the framework.  Additional content experts then review and refine the resulting outline until consensus is reached that the key elements of the problem have been accounted for.  This expert-guided scheme then serves as a preliminary coding structure for analysis of student work.  If necessary, the operationalization can be further refined to accommodate aspects of student problem-solving that were not captured by the expert task analysis.

In prior work, we operationalized the ACER framework for the use of multi-variable integration in the context of calculating the electric potential from a continuous charge distribution \cite{wilcox2013acer}.  Here, we modify this operationalization such that it is appropriate for the use of multi-variable integration to calculate potentials in a wider range of contexts including gravitation and magnetostatics.  The elements of the modified operationalized ACER framework are detailed below.  Element codes are for labeling purposes only and are not meant to suggest a particular order, nor are all elements always necessary for every problem.  In particular, the elements of Construction and Execution are unlikely to occur in the specific order listed as experts can, and often do, iterate back and forth between setting up and evaluating expressions.

\textbf{Activation of the tool}: The first component of the framework involves the selection of a solution method.  The modified task analysis identified four elements that are involved in the activation of resources identifying multi-variable integration as the appropriate tool.

\vspace{-2mm}
\begin{enumerate}[label={\bf A\arabic*}:, align=left] \itemsep-10pt \parskip0pt \parsep0pt
   \item The problem asks for a vector field or the associated potential.\\
   \item The problem provides an expression for, or description of, the distribution that is the source of the field or potential.\\
   \item The provided distribution does not have appropriate symmetry to utilize mathematically simpler approaches (e.g., variations of Gauss' Law).\\
   \item Direct calculation of the potential is more efficient than direct calculation of the associated field.
\end{enumerate}
\vspace{-2mm}

Elements {\bf A1--A3} are cues typically present in the problem statement. Element {\bf A4} is specific to problems asking for the potential and is included to account for the possibility of solving for potential by first calculating the associated field.  This method is valid but often more difficult.

\textbf{Construction of the model}: Here, mathematical resources are used to map the specific physical situation onto the general integral expression.  The integral expression produced at the end of the \emph{construction} component should be in a form that could, in principle, be solved with no knowledge of the physics of this specific problem.  We identify four key elements that must be completed in this mapping.

\vspace{-2mm}
\begin{enumerate}[label={\bf C\arabic*}:, align=left] \itemsep-10pt \parskip0pt \parsep0pt
   \item Use the geometry of the distribution to select a coordinate system. \\
   \item Express the differential element (e.g., $dm$ or $\vec{J}dV'$) in the selected coordinates. \\
   \item Select integration limits consistent with the differential charge element and the extent of the physical system. \\
   \item Express the difference vector, $|\vec{r}-\vec{r}'|$, in the selected coordinates. \\
\end{enumerate}
\vspace{-2mm}

Elements {\bf C2} and {\bf C4} can be accomplished in multiple ways, often involving several smaller steps.  In order to express the differential element, the student must combine the charge density and differential to produce an expression with dimensions consistent with the physical quantity being summed over (e.g., $\vec{J} dV$).  Construction of the difference vector often includes a diagram that identifies both the source, $\vec{r}'$, and field point, $\vec{r}$, vectors.

\textbf{Execution of the Mathematics}: This component of the framework deals with the mathematics required to compute a final expression.  In order to produce a formula describing the potential or field, it is necessary to:

\vspace{-2mm}
\begin{enumerate}[label={\bf E\arabic*}:, align=left] \itemsep-10pt \parskip0pt \parsep0pt
   \item Maintain an awareness of which variables are being integrated over (e.g., $r'$ vs. $r$). \\
   \item Execute (potentially multi-variable) integrals in the selected coordinate system. \\
   \item Manipulate the resulting algebraic expressions into a form that can be readily interpreted. \\
\end{enumerate}
\vspace{-2mm}

\textbf{Reflection on the result}: The final component of the framework involves verifying that the expression is consistent with expectations.  While many different techniques can be used to reflect on the result, these two checks are particularly common for problems involving multi-variable integration:

\vspace{-2mm}
\begin{enumerate}[label={\bf R\arabic*}:, align=left] \itemsep-10pt \parskip0pt \parsep0pt
   \item Check the units of intermediate and final expressions. \\
   \item Check the limiting behavior to ensure it is consistent with the nature and geometry of the distribution. \\
\end{enumerate}
\vspace{-2mm}

Element {\bf R2} is especially useful when the student already has some intuition for how the potential or field should behave in particular limits.  However, if they do not come in with this intuition, reflection on the results of this type of problem is a vital part of developing it.

In the next section, we will apply this operationalization of ACER to investigate and compare students' reasoning and prevalent difficulties when solving the physics problems described in the previous section (Sec.\ \ref{sec:data}) involving integration in the context of gravitation, electrostatics, and magnetostatics.

\section{\label{sec:results}Results}

Here we present results from our investigations of students' use of direct integration.  Since our goal is to compare changes in students' reasoning across multiple points in the curriculum, we organize our results here by component of the ACER framework and present data from the context of gravitation, electrostatics, and magnetostatics together.  Throughout this section, we also compare to our prior work investigating students' use of integration in the context of electrostatics to provide an additional data point to understand differences in students' reasoning across multiple contexts.  This earlier work was described in Sec.\ \ref{sec:acer} and also involved analysis of students' exam solutions and interview responses to questions similar to those considered here for the scalar potential, but involving calculating the electric potential from a line, surface, and volume charge distributions.

\subsection{\label{sec:activation}Activation of the tool}

One of the two vector potential exam question in the E\&M course provided students with Eqn.\ \ref{eqn:a}, effectively bypassing the \emph{activation} component of the problem-solving process.  Here, we will focus on data from the remaining courses, representing $N=325$ exam solutions.  Evidence for the individual elements of \emph{activation} within the exam solutions can be difficult to identify because student often do not articulate their thought process on the exams.  In particular, there was rarely explicit evidence that students attended to elements A1 (i.e., the prompt asked for the potential) or A2 (i.e., the prompt provided a distribution).  These elements, while a critical part of fully justifying direct integration as the appropriate tool, are often a tacit step in the problem-solving process.

Elements A3 (i.e., recognizing the distribution does not have sufficient symmetry) and A4 (i.e., recognizing direct calculation of the potential is most efficient) were most easily identified when students did not approach the problem using the appropriate mathematical tool (see Table \ref{tab:activation} for a summary).  For example, students who attempt to calculate the potential by first calculating the associated field or force by treating the distribution as highly symmetric (e.g., treating it as a point source or using Gauss's Law) have demonstrated a difficulty with element A3.  This difficulty was manifested by roughly a tenth of the solutions in the contexts of both gravitation (7\%, $N=5$ of 77) and the scalar potential (14\%, $N=23$ of 160).  The increase in the fraction of students exhibiting difficulties with element A3 between Class.\ Mech.\ and E\&M was driven in part by students' using Gauss's Law to calculate the electric field.  Difficulty with element A3 was observed in none of the $N=50$ vector potential solutions, possibly because no ``point-source'' approximation exists for the vector potential due to the lack of magnetic monopoles.  However, we also observed no student attempting to approach the rod of current question using Ampere's Law.   

Another observed difficulty with \emph{activation} was attempting to calculate the potential by first calculating the associated field or force via direct integration and then taking the line integral or, in the case of gravitation, using $U=mgh$.  The former approach is valid in the case of gravitation and the scalar potential, though significantly more difficult, and represents a difficulty with element A4.  The latter approach represents a difficulty with both A4 and A3, where the student has taken a harder route by opting to calculate the gravitational field $\vec{g}$, and also over generalized the equation $U=mgh$, which is only valid near the surface of the earth.  Students with difficulties with element A4 represented just over 10\% of students in the context of both gravitation (12\%, $N=9$ of 77) and the scalar potential (14\%, $N=23$ of 160).  In the case of the vector potential, the approach of first calculating the magnetic field is not a valid approach as the relation between the vector potential and magnetic field is more complex (i.e., $\vec{B}=\nabla \times \vec{A}$), and only one of the $N=50$ vector potential solutions attempted this approach.

\begin{table*}
\caption{Difficulties activating direct integration via Eqns.\ \ref{eqn:u}-\ref{eqn:a} as the appropriate mathematical tool for exam questions featuring a rod or strip distribution (e.g., Fig.\ \ref{fig:equestions}).  Percentages are given with respect to the subset of students with difficulties with \emph{activation} (given in parentheses).  Codes are not exhaustive or exclusive (so percents may not total to 100\%) but represent the most common themes.}\label{tab:activation}
 \begin{ruledtabular}
    \begin{tabular} {c l c c c c c c}
	 &  &	\multicolumn{2}{l}{{\bf Gravitation}} & \multicolumn{2}{l}{{\bf Electrostatics}}   & \multicolumn{2}{l}{{\bf Magnetostatics}}\\
	{\bf Element} & \textbf{Difficulty} 					& \textbf{N} 	& \textbf{\%} & \textbf{N} & \textbf{\%} & \textbf{N} & \textbf{\%}\\
	\hline
	A3 & Point source or Gauss								&  5	& 33\% (of $N=15$) & 23 & 49\% (of $N=47$) & 0 & 0\% (of $N=1$)	\\
	A4 & Calculate of force or field 		&  9	& 60\% (of $N=15$) & 23 & 49\% (of $N=47$) & 1 & 100\% (of $N=1$)	\\
    \end{tabular}
  \end{ruledtabular}
\end{table*}

Overall, roughly three-quarters of students (80\%, $N=62$ of 77 in gravitation; 70\%, $N=113$ of 160 in electrostatics) correctly activated direct integration via Eqns.\ \ref{eqn:u}-\ref{eqn:v} as the correct approach to the exam questions (Fig.~\ref{fig:equestions}).  These results are consistent with what was found in our earlier investigation in electrostatics where just under three quarters (73\% \cite{wilcox2013acer}) of students correctly used Eqn.~\ref{eqn:v} to calculate the electric potential in problems involving various charge distributions with the most common alternative approach being to attempt the problem by first calculating the electric field via either Coulomb's law or Gauss's law.  This suggests that, while Eqn.~\ref{eqn:v} is at least the second time in the curriculum that students have encountered the direct integration approach, they are not showing a significant increase in success rates for \emph{activation} of this tool, and potentially showing some increased difficulties.  In both of these cases, students typically enter the course already familiar with the potential or vector field in question in the context of simple and symmetric distributions.  Class.\ Mech.\ students have seen the gravitational field from a point mass or near the surface of the earth (where $U=mgh$ is a valid expression), and E\&M students have calculated the electric field from highly symmetric charge distributions using Gauss's Law. 

Together these results suggest in all contexts, the challenge for students with respect to \emph{activation} may be eliminating inappropriate but simpler and/or more familiar methods as viable solution paths.  This difficulty appears in 20-30\% of solutions in both contexts where such simpler/more familiar methods exists, though the pseudo-longitudinal nature of our data make it impossible to determine if this is because the students who struggle in the context of gravitation continue to struggle later in electrostatics, or if some students who did not encounter this difficulty in gravitation later encounter it in electrostatics and \emph{vice versa}.

The interviews provide additional insight into the issue of \emph{activation}.  Neither the gravitation nor the vector potential interview questions prompted students to utilize Eqns.~\ref{eqn:u} or \ref{eqn:a}, and difficulties with elements A3 and A4 were observed in both the sets of interviews.  One of the four pairs of gravitation interviewees treated the given ring (see Fig.~\ref{fig:iquestion}) as a point mass, representing a failure in A3.  One student argued that any rigid body could be treated as a point at its center of mass; the other was less confident in this approach, but said that they did not know how to account for the details of this particular mass distribution.  When prompted to consider approaching the problem via direct integration, both students expressed concerns about the potential for ``horrible algebra.''

These latter two sentiments also arose in the vector potential interviews wherein two of three pairs attempted to calculate the magnetic field via Ampere's law, planning to use the result to ``solve'' for the vector potential by inverting the relationship $\vec{\nabla} \times \vec{A} = \vec{B}$ (i.e., a difficulty with element A3).  This approach is ultimately not possible, and both pairs struggled to find a useful Amperian loop. When directly prompted by the interviewer to consider Eqn.~\ref{eqn:a}, both pairs struggled to use it (discussed in Sec.~\ref{sec:construction}), and subsequently returned to the Ampere's law strategy despite their earlier trouble.  This difficulty, while not observed in the exam solutions, is consistent with our hypothesis that difficulties with \emph{activation} are linked primarily to eliminating simpler or more familiar approaches.  The interviews provide additional detail to this interpretation by suggesting that both perceived and actual difficulties with the direct integration approach may impede activation of this tool and make students less likely to eliminate what they perceive as easier approaches even if they doubt their applicability.

The choice to calculate the magnetic field via Ampere's law reflects difficulties with element A3.  None of the vector potential interview students attempted to calculate the magnetic field via direct integration as a step toward finding the vector potential (i.e., element A4), though one pair spontaneously considered it, but preferred not to attempt it.  Alternatively, in the gravitation context, difficulties with element A4 were observed independent of difficulty with A3; two of the four pairs of gravitation interviewees calculated the gravitational force instead of potential energy.  This includes the pair who first answered with the point mass expression for potential energy, but, when prompted by the interviewer to attempt integration, switched to calculating the force.  However, only one of these two pairs could name the correct relationship between force and potential energy that would allow them to answer the question fully.

\subsection{\label{sec:construction}Construction}

\begin{table*}
\caption{Difficulties expressing the differential element (i.e., $dm$, $dq$, or $JdV'$).  Percentages are given with respect to the subset of students with difficulties with element C2 (given in parentheses).  Codes are not exhaustive or exclusive (so percents may not total to 100\%) but represent the most common themes.}\label{tab:construction2}
  \begin{ruledtabular}
    \begin{tabular} {l c c c c c c}
	   &	\multicolumn{2}{l}{{\bf Gravitation}} & \multicolumn{2}{l}{{\bf Electrostatics}}   & \multicolumn{2}{l}{{\bf Magnetostatics}}\\
	 \textbf{Difficulty} 								& \textbf{N} 	& \textbf{\%} & \textbf{N} & \textbf{\%} & \textbf{N} & \textbf{\%}\\
	\hline
	Incorrect differential line or area element 			& 10 & 83\% (of $N=12$) & 11 & 61\% (of $N=18$) & 21 & 51\% (of $N=41$)	\\
           \hspace{3mm} e.g., $dl=dx$ rather than $dx'$   	&   	& 			&   	&			&		&	\\
           \hspace{10mm} or $dl=da$  						& 		& 			&   	&			&		&	\\
	Incorrect expression for density						& 5 & 42\% (of $N=12$) & 8 & 44\% (of $N=18$) & 27 & 66\% (of $N=41$)	\\
           \hspace{3mm} e.g., $\lambda = M$ or $I=J$  	&   	& 			&   	&			&		&	\\
    \end{tabular}
  \end{ruledtabular}
\end{table*}

In all three contexts, the largest number of student difficulties were observed in the \emph{construction} component where students are operationalizing the general integral forms in Eqns.\ \ref{eqn:u}-\ref{eqn:a} for the specific physical situation given in the problem.  Overall, $N=72$ (out of 77) of the gravitation solutions, $N=130$ (out of 160) of the scalar potential solutions, and $N=135$ (out of 138) of the vector potential solutions included at least one element of \emph{construction} in their solution (i.e., they used direct integration and did more than just write an equation or final answer).  With respect to element C1 (i.e., selecting an appropriate coordinate system), all exam prompts provided a figure which included a pre-selected origin and labeled Cartesian axes (effectively bypassing this element), and only 3 students in the entire data set utilized anything other than Cartesian variables.  Similarly, all but one pair of interviewees appropriately chose polar coordinates, with the final pair opting for Cartesian coordinates.  One E\&M interviewee did however express concerns about the use of curvilinear coordinates, worried that the inseparability of Poisson's equation for $\vec{A}$ in this system may also invalidate Eqn.~\ref{eqn:a}.

Element C2 proved more challenging for students with roughly a fifth of the gravitation solutions (17\%, $N=12$ of 72), just over a tenth of the scalar potential solutions (14\%, $N=18$ of 130), and a third of the vector potential solutions (30\%, $N=41$ of 135) providing responses reflecting difficulties expressing the differential element (see Table~\ref{tab:construction2}).  For the questions considered here, these difficulties could manifest in two primary ways: difficulties expressing the density (e.g., $\lambda$, $\vec{J}$) or difficulties expressing the differential line or area element ($dl$ or $dA$).  

For the students in the context of gravitation, difficulty expressing the differential line element was most common (83\%, $N=10$ of 12) and included expressing $dl$ as a function of the field variables (i.e., $r$ vs. $r'$, $N=5$) and expressing the differential element as an area or volume element ($N=2$).  Similar difficulties with the differential line element were observed in the E\&M students' solutions, though represented a smaller fraction of the difficulties with element C2 (61\%, $N=11$ of 18 in electrostatics; 51\%, $N=21$ of 41 in magnetostatics).  Moreover, focusing only on the exam questions featuring line distributions, the overall fraction of students exibiting difficulties with the differential line element decreases from 14\% ($N=10$ of 72) to 9\% ($N=20$ of 218) from Class.\ Mech.\ to E\&M.  This downward trend across consecutive exposures suggests that difficulty with element C2 may be one that fades over time as students gain more experience. 

None of the interviewed Class.\ Mech.\ pairs had trouble writing the differential line element in polar coordinate.  It is possible that some of these students may have had more difficulty on their own --- one, for example, first suggested $dl=Rd\theta d\phi$ --- but as pairs, the students were able to correct each others mistakes.  The E\&M interviewees were equally successful, with the exception of the pair working in Cartesian coordinates, who wrote $dl=\sqrt{dx^2 + dy^2}$.  However, this pair did generate the correct form once prompted to switch to polar coordinates.  

Difficulties expressing the linear mass density $\lambda$ were less common (42\%, $N=5$ of 12) for students the context of gravitation and limited almost exclusively ($N=4$) to incorrectly substituting the total mass of the rod for $\lambda$.  This difficulty was observed with similar frequency in the context of the scalar potential (44\%, $N=8$ of 18).  At first glance, this trend would also seem to suggest that students difficulties around the density stay relatively constant over time; however, this pattern does not continue in the context of magnetostatics.  One version of the vector potential exam question provided students with the expression for the surface current density $\vec{K}$ (see Fig.~\ref{fig:em1}), effectively bypassing the need for students to express this quantity.  Alternatively, the other version of the question required students to express (or translate) the volume current density $\vec{J}$ for the line of current on their own.  In this case, difficulties around the current density represented the majority of the difficulties with element C2 (90\%, $N=27$ of 30 solutions exhibiting difficulties with C2).  This difficulty most often (63\%, $N=17$ of 27) manifested as students simply leaving $J$ in their final expression without transitioning to the current $I$ carried by the wire segment (i.e., incorrectly replacing $JdV' \rightarrow Jdz'$) and resulting in a final expression with the wrong units.

Consistent with the results of the exam coding, no issues with the linear mass density arose in the Class.\ Mech.\ interviews, but all three E\&M pairs experienced some degree of difficulty with the current density.  Two pairs did, after unguided discussion, generate a valid expression for the magnitude of $\vec{J}$ using delta-functions, but both omitted the direction (although one pair caught this error much later on).  Both pairs used the units of current density to guide their choice of delta-functions.  The third pair also referenced units, but argued that the units of $\vec{J}$ were current per volume.  They were unable to find an expression for the current density and were eventually provided with the one-dimensional form of Eqn.~\ref{eqn:a}.

One interpretation of this sudden increase in the frequency of a difficulty that otherwise seemed to be holding steady as students progressed may be that it is, in fact, a totally new difficulty.  The volume current density, $\vec{J}$, is a conceptually and mathematically more challenging quantity than either mass or charge density.  Moreover, while the quantity $\rho dV'$ (where $\rho$ is volume mass density) has a clear physical interpretation as the total mass on each differential chunk of the object (i.e., $dm$), the quantity $\vec{J}dV'$ has no such clear physical interpretation.  One exchange between E\&M interviewees discussing the meaning of the integral in Eqn.~\ref{eqn:a} demonstrates this difficulty.  First, they associated ``the current'' with the entire ring of current, making the choice of a specific $\vec{r}'$ which represents the entire circle ambiguous.  One student then had the insight that $\vec{r}'$ points to ``a current density at a given location along the loop,'' and used the phrase ``for each J'' to describe this.  Their partner objected to this language as there is only one function $\vec{J}$, and asks if they meant ``differential $J$.''  While over the course of this exchange the pair made progress towards understanding Eqn.~\ref{eqn:a}, their interpretation of the differential element was still dimensionally incorrect (they had not accounted for the volume element).  Moreover, their difficulties were exacerbated by the lack of any common terminology with which to refer to the quantity being summed over.

After constructing an expression for the differential element, students must select limits of integration that are consistent with this element and the physical extent of the system (element C3).  Difficulties with the limits appeared in less than a fifth of the gravitation solutions (19\%, $N=14$ of 72), roughly a quarter of the scalar potential solutions (28\%, $N=36$ of 130), and roughly a tenth of the vector potential solutions (12\%, $N=16$ of 135).  In all cases, the two most common issues were not integrating over the actual extent of the rod ($N=11$ of 14 in gravitation, $N=26$ of 36 in electrostatics, and $N=7$ of 16 in magnetostatics), and not including limits at all ($N=1$ of 14 in gravitation, $N=8$ of 36 in electrostatics, and $N=9$ of 16 in magnetostatics).  Combined with findings from our previous work in the context of electrostatics showing difficulties with limits in only 14\% of solutions \cite{wilcox2013acer}, these data suggest that difficulties with element C3 may fluctuate somewhat over students first exposures to direct integration in these contexts, but stay present to at least some extent even after multiple exposures.  In the interviews, none of the six pairs who made it to this point in the interview struggled to choose limits of $0$ to $2\pi$ for the ring.  This includes the pair working in Cartesian coordinates who still articulated this choice, suggesting that for some the limits may have been an automatic response to the circular geometry rather than a considered decision.

The fourth element in \emph{construction} (C4) deals with the difference vector ($\vec{r}-\vec{r}'$) which points from the source point to the field point.  In our prior investigation, nearly half of students encountered difficulties when attempting to express this vector \cite{wilcox2013acer} for a disk-shaped charge distribution.  The distributions provided in this study results in a simpler expression for difference vector, and, unsurprisingly, students had somewhat less difficulty producing a correct expression with 30-45\% of students unable to do so (30\%, $N=22$ of 72 in gravitation; 45\%, $N=45$ of 130 in electrostatics; 35\%, $N=47$ of 135 in magnetostatics).  However, the fact that the fraction of students exhibiting difficulties with the difference vector remains high, and perhaps even increases slightly, suggests that this difficulty is persistent across multiple exposures. 

Lending detail to the results from the exams, two types of difficulties related to the difference vector arose in the interviews.  In two of the four Class.\ Mech.\ interviews, the students generated incorrect expressions for the magnitude of the difference vector.  When prompted to consider more carefully or to draw the relevant vectors, however, both pairs came to a valid expression.  These students appeared to understand the meaning of the difference vector term in the integrand (i.e., the distance from source to test), but had made geometric errors in calculating its magnitude.  Alternatively, some of the E\&M pairs struggled with the meaning of the $\vec{r}-\vec{r}'$ term.  This difficulty appeared to be related to Griffiths' ``script-$r$'' notation (script-$r = \vec{r} - \vec{r}'$) \cite{griffiths1999em}).  A total of four pairs of students used this notation either spontaneously or when prompted.  At least one student in each pair expressed doubts about the relationship between script-$r$, $\vec{r}$, and $\vec{r}'$ or the meaning of each vector (e.g., switching the primed and un-primed variables).  Two of these pairs had sufficient difficulty remembering or generating these ideas that they never attempted to write an expression for the difference vector or its magnitude.   While the convention of using script-$r$, $\vec{r}$, and $\vec{r}'$ permits concise formulas, it requires students to parse — or, in the case of the interviewees in this study, simply try to remember — the meaning of these symbols before they can proceed to analyzing the geometry of the problem at hand.  This finding is consistent with our previous findings in the context of electrostatics \cite{wilcox2013acer}.

\subsection{\label{sec:execution}Execution}

The Execution component of the framework deals with the procedural aspects of working with mathematical tools in physics.  Overall, $N=66$ of the gravitation solutions, $N=128$ of the electrostatics solutions, and $N=77$ of the magnetostatics solutions included at least one element of \emph{execution}.  On one of the vector potential exam questions, students were only asked to set up the integral and thus did not include \emph{execution}.  In the remaining exam questions, it is not possible to know for certain from an exam solution whether students keep track of which variables are being integrated over (i.e., source vs. field variables - element E1); however, we can identify cases where students explicitly integrated over the wrong variables (i.e., integrating over the field variables).  Roughly a tenth of the students in both courses ($N=5$ of 66 in gravitation, $N=20$ of 128 in electrostatics, and $N=7$ of 77 in magnetostatics) explicitly integrated over variables in their expression which represented the field point. 

Only the scalar potential questions resulted in integrals that could be reasonably calculated by hand, and just over a third (37\%, $N=47$ of 128) of solutions exhibited one or more mistakes in doing so.  The other questions resulted in integrals for which the general solution was provided to the students (effectively bypassing element E2).  However, some students still attempted to perform integrals, typically because a mistake in an earlier step of the solution resulted in a solvable integral.  Of the $N=23$ Class.\ Mech.\ students who attempted to perform integrals just under half (43\%, $N=10$ of 23) made significant errors in the process.  This fraction dropped to only 17\% ($N=3$ of 18) in the context of the vector potential students who attempted to perform an integral.  Note that these students had, by necessity, already made one or more errors in the \emph{activation} and \emph{construction} components of the framework before reaching \emph{execution}.


Given the high pressure and individual nature of both exams, we expected that many students would make mathematical errors particularly with element E3 (algebraic manipulation).  To account for this, we distinguish in our analysis between small math errors (e.g., dropping a factor of 2 or minus sign) and more fundamental mathematical errors (e.g., dropping the bottom limit or executing integrals incorrectly).  In the context of the scalar potential, just over half of the students who made mathematical errors made more fundamental errors (60\%, $N=28$ of 47); however, for roughly three quarters of these students (71\%, $N=20$ of 28), the mistake made in the \emph{execution} stage was preceded by a significant difficulty in the \emph{activation} or \emph{construction} components.  This trend is consistent with the findings of our prior work that suggested difficulty performing integrals was rarely the primary barrier to students' success when using integrals in the context of physics problems \cite{wilcox2013acer}.  


In interviews, students were not asked to evaluate the integral expression they derived in the interviews, although one pair did switch from integrating over the source variables to over the field variables when considering how they might do so.  Otherwise the interviews provided little insight into the \emph{execution} component.

\subsection{\label{sec:reflection}Reflection}

The reflection component deals with the process of checking and/or interpreting the final expression. It is often the case that mistakes in the \emph{construction} or \emph{execution} components resulted in an expression for the potential that had the wrong units or did not have the correct behavior in particular limits (i.e., elements R1 and R2, respectively). Overall, we found that very few of our students ($N = 20$ of the 329 students in any of the three contexts to complete the question) made explicit, spontaneous attempts to reflect on their solution using either of these checks on exams. This number should be interpreted as a lower bound on the frequency of spontaneous reflections, as it is possible that more of the exam students made one of these checks and simply did not write it down explicitly on their exam solution; however, the interview results suggest this is less likely.  Although all three E\&M interview pairs considered units when writing $\vec{J}$, and one pair checked extreme cases for their difference vector expression, none of the interviewees spontaneously reflected on their final integral expression.  Some explained that they used these tools only when particularly worried about a result, and only if they felt they had time.

Another strategy for understanding reflection involves looking at the number of solutions where the final expression included an error that would have been detected by one or more of these checks. Table \ref{tab:reflection} lists this along with the number of solutions that explicitly included each reflective check. Overall, these results suggest that an explicit check of units would likely be the most effective reflective practice for students in terms of detecting errors, but that our students are rarely executing this (or other) checks spontaneously.  This finding is consistent with our prior investigation \cite{wilcox2013acer} and implies that students do not grow more likely to spontaneously execute these kinds of reflective behaviors as they progress further through the curriculum. 

\begin{table}
\caption{Number of exam students who explicitly utilized each of the two common reflective checks ($N_{explicit}$) along with the number of solutions that included an error that would have been detected by this check ($N_{incorrect}$).  Data from all three contexts have been combined here due to low $N_{explicit}$. The total number of students who progressed far enough in their solution to utilize one of these checks was $N=329$.  }\label{tab:reflection}
  \begin{ruledtabular}
    \begin{tabular} {l c c}
	\textbf{Reflective check} & \textbf{$N_{incorrect}$} 	& \textbf{$N_{explicit}$}\\
	\hline
	Units (R1)		& 78	&  3	\\
	Limits (R2) 	& 62 	&  10  	\\
    \end{tabular}
  \end{ruledtabular}
\end{table}

We also investigated whether students can perform these reflective checks when asked to do so.  In one of the two scalar potential exam questions, two followup questions asked students to discuss how the potential should behave as you got far from the charge distribution and to confirm that their expression was consistent with this expectation.  Most students (72\%, $N=47$ of 65) correctly articulated that the potential should fall off as $1/r$ in the limit of large-$r$.  The most common alternative answer was simply that the potential ``goes to zero" with no discussion of how it goes to zero (17\%, $N=11$ of 65).  When asked to show that their answer matched this behavior, nearly half (44\%, $N=24$) of the 55 students who responded executed an appropriate Taylor expansion of their prior solution, and a further quarter (25\%, $N=14$) had made prior mistakes in their solution such that their solution already had a pure $1/r$ dependence.  The remaining quarter of students (24\%, $N=13$) simply plugged in $r= \infty$ into their expression and showed that their solution went to zero for large-$r$. 

The interviews also provided insight into students prompted reflection.  Five of the seven interviews included a second question targeted at \emph{reflection}.  All five pairs suggested and were able to carry out a units check, although only one pair did so without first expanding the relevant constant ($G$ or $\mu_0$) in fundamental units.  All five also checked that the expressions vanished in the limit $r \to \infty$.  On the other hand, three pairs expected the expressions to blow up in the $r \to 0$ limit, but this criterion incorrectly eliminated one of the proposed answers.  Moreover, only one pair successfully determined the large-$r$ functional form of this answer, and only when prompted to do so.  

Interviewees were also asked to reflect on their final answers for the potential from the ring given in the first interview question.  Their strategies here matched those employed in the second question, but it is worth noting that two students expressed uncertainty as to the effect of the integral on units or limits.  Combined with the exam results, this suggests that students are capable of checking units and limiting values (e.g., that a function goes to zero) but are both less inclined towards, and have more difficulty, checking limiting forms (e.g., \emph{how} the potential goes to zero).

\section{\label{sec:discussion}Summary and Implications}

Here, we build on prior work investigating students' use of direct integration as a mathematical tool in physics problem solving.  We extend this prior work, which focused on students difficulties in the context of junior-level electrostatics, by investigating students use of integration in two additional points in the undergraduate curriculum: in the context of gravitation at the sophomore level, and magnetostatics at the junior-level.  With the goal of making pseudo-longitudinal comparisons across these three different content areas, we again utilized the ACER framework to structure the design and analysis of our investigations.  This has allowed us to directly compare the reasoning and difficulties students exhibit at key points in the problem-solving process when utilizing direct integration, and determine whether and how these difficulties evolve or shift as students encounter the same mathematical tool multiple times in different contexts.  

With respect to \emph{activation}, we found that across all three contexts some students (roughly a quarter) showed difficulties in identifying direct integration as the correct approach.  In nearly all cases, students who did not use direct integration for the potential instead tried to approach the problem by first calculating the associated field by another method.  Moreover, in interviews, students often expressed concern about direct integration being too challenging, instead opting for more familiar approaches or approaches perceived as being easier.  This tendency to avoid particular approaches because they are ``too hard" may well be something these students have learned implicitly their courses.  Physicists tend to give students questions that can be solved in a relatively straight-forward manner and often employ tricks to simplify the mathematics of a problem.  This tendency may have the unintended consequence of making students less willing to attempt approaches they consider to be mathematically complex in the belief that there must be an easier approach.  Moreover, this difficulty does not appear to fade significantly over multiple exposures.  This suggests that instructors may need to place greater emphasis on explicit discussions of how to identify the correct approach to a problem and how to eliminate mathematically simpler approaches when they are not applicable.  

In terms of \emph{construction}, we found a greater degree of variation in students' difficulties in the different contexts.  In all contexts, one of the primary \emph{construction} difficulties encountered related to expressing the differential element; however, the nature of this difficulty varied significantly.  Difficulties expressing the differential line element were most prevalent in the context of gravitation, and gradually became less frequent in the context of electrostatics and then magnetostatics.  This may suggest that this difficulty fades somewhat over the course of the current curriculum.  Alternatively, the primary difficulties in expressing the differential element in the context of magnetostatics related to the current density, while difficulties with the mass or charge density were less common in the context of gravitation and electrostatics and occurred with roughly the same frequency.  We argue that this difficulty, while manifest in the mathematics of the problem, actually reflects a conceptual difficulty relating to interpreting the current density itself and the physical meaning of the $\vec{J}dV'$ term.  This represents an example of a situation where a new difficulty not observed in early exposures to direct integration arose specifically as a consequence of a change in the physical context of the problem.  One other element of \emph{construction} which saw significant difficulties was expressing the difference vector.  However, the frequency of this difficulty stayed more consistent across the context of three contexts, and perhaps even got worse after the introduction of ``script-r,'' with between a third and a half of students displaying difficulties in both cases, suggesting that this difficulty is also persistent over multiple exposures.  

Based on the findings of the earlier study of students' use of direct integration in electrostatics, which found that difficulties around the \emph{execution} were rarely the primary barrier to student success, the current study included only a few questions asking students to work through the mechanics of actually performing integrations.  However, consistent with this previous finding, the majority of \emph{execution} errors observed in this study were made by students who had already made one or more significant mistakes in the \emph{activation} or \emph{construction} components of the their solution.  This result, once again, suggests that the procedural mistakes made during the integration process were not the most significant issues encountered by these students.  

Finally, with respect to \emph{reflection}, we found that in all three contexts spontaneous bids for \emph{reflection} were very rare in both the exams and interviews.  This suggests that multiple exposures to direct integration in physics problem solving do not appear to encourage students to execute reflective checks on their solutions when not prompted to do so.  Exams and interviews further suggested when prompted to come up with possible reflective checks, students in all three contexts are able to identify appropriate possibilities (e.g., checking units or limits).  However, interviews also suggested that actually executing these checks, particularly in cases where doing so requires extra steps such as executing a Taylor expansion, may be more challenging for students, particularly after only early exposure to direct integration in the context of gravitation.  

Overall, the results of this study, combined with our earlier investigation, provide insight into students' use of direct integration in multiple contexts and provides examples of cases where observed students difficulties get better, change, and remain the same over the course of these multiple exposures.  The results can be used to help instructors identify areas where additional effort is needed to address persistent issues (e.g., identifying integration as the appropriate tool, and expressing the difference vector), versus areas where difficulties fade with experience (e.g., expressing differential line, area, and volume elements), as well as to anticipate new difficulties that appear in different contexts (e.g., interpreting and expressing the current density).  


This work represents a novel example of how the ACER framework can serve as a standardized tool to facilitate cross-study comparisons of students' use of mathematical tools in physics problem solving.  Future work will involve cross-context comparisons of students use of other mathematical tools in other physics contexts.  For example, building on prior work with the Dirac delta function in the context of electrostatics could be extended to include more mathematically complex uses in quantum mechanics or Fourier transforms.  Similarly, investigations of students' use of separation of variables in electrostatics could form the basis of understanding changes in students' difficulties when they encounter this tool again in quantum mechanics.

\begin{acknowledgments}
This work was supported by the University of Colorado Physics department.  Particular thanks to the members of PER@C for all their help and feedback, as well as the students and faculty who participated in the study.
\end{acknowledgments}

\bibliography{Integration-refs}

\begin{thebibliography}{20}%
\makeatletter
\providecommand \@ifxundefined [1]{%
 \@ifx{#1\undefined}
}%
\providecommand \@ifnum [1]{%
 \ifnum #1\expandafter \@firstoftwo
 \else \expandafter \@secondoftwo
 \fi
}%
\providecommand \@ifx [1]{%
 \ifx #1\expandafter \@firstoftwo
 \else \expandafter \@secondoftwo
 \fi
}%
\providecommand \natexlab [1]{#1}%
\providecommand \enquote  [1]{``#1''}%
\providecommand \bibnamefont  [1]{#1}%
\providecommand \bibfnamefont [1]{#1}%
\providecommand \citenamefont [1]{#1}%
\providecommand \href@noop [0]{\@secondoftwo}%
\providecommand \href [0]{\begingroup \@sanitize@url \@href}%
\providecommand \@href[1]{\@@startlink{#1}\@@href}%
\providecommand \@@href[1]{\endgroup#1\@@endlink}%
\providecommand \@sanitize@url [0]{\catcode `\\12\catcode `\$12\catcode
  `\&12\catcode `\#12\catcode `\^12\catcode `\_12\catcode `\%12\relax}%
\providecommand \@@startlink[1]{}%
\providecommand \@@endlink[0]{}%
\providecommand \url  [0]{\begingroup\@sanitize@url \@url }%
\providecommand \@url [1]{\endgroup\@href {#1}{\urlprefix }}%
\providecommand \urlprefix  [0]{URL }%
\providecommand \Eprint [0]{\href }%
\providecommand \doibase [0]{http://dx.doi.org/}%
\providecommand \selectlanguage [0]{\@gobble}%
\providecommand \bibinfo  [0]{\@secondoftwo}%
\providecommand \bibfield  [0]{\@secondoftwo}%
\providecommand \translation [1]{[#1]}%
\providecommand \BibitemOpen [0]{}%
\providecommand \bibitemStop [0]{}%
\providecommand \bibitemNoStop [0]{.\EOS\space}%
\providecommand \EOS [0]{\spacefactor3000\relax}%
\providecommand \BibitemShut  [1]{\csname bibitem#1\endcsname}%
\let\auto@bib@innerbib\@empty
\bibitem [{\citenamefont {Hsu}\ \emph {et~al.}(2004)\citenamefont {Hsu},
  \citenamefont {Brewe}, \citenamefont {Foster},\ and\ \citenamefont
  {Harper}}]{hsu2004problemsolving}%
  \BibitemOpen
  \bibfield  {author} {\bibinfo {author} {\bibfnamefont {Leon}\ \bibnamefont
  {Hsu}}, \bibinfo {author} {\bibfnamefont {Eric}\ \bibnamefont {Brewe}},
  \bibinfo {author} {\bibfnamefont {Thomas}\ \bibnamefont {Foster}}, \ and\
  \bibinfo {author} {\bibfnamefont {Kathleen}\ \bibnamefont {Harper}},\
  }\bibfield  {title} {\enquote {\bibinfo {title} {Resource letter rps-1:
  Research in problem solving},}\ }\href@noop {} {\bibfield  {journal}
  {\bibinfo  {journal} {Am. J. Phys.}\ }\textbf {\bibinfo {volume} {72}},\
  \bibinfo {pages} {1147--1156} (\bibinfo {year} {2004})}\BibitemShut {NoStop}%
\bibitem [{\citenamefont {Meltzer}\ and\ \citenamefont
  {Thornton}(2012)}]{meltzer2012resource}%
  \BibitemOpen
  \bibfield  {author} {\bibinfo {author} {\bibfnamefont {David~E}\ \bibnamefont
  {Meltzer}}\ and\ \bibinfo {author} {\bibfnamefont {Ronald~K}\ \bibnamefont
  {Thornton}},\ }\bibfield  {title} {\enquote {\bibinfo {title} {Resource
  letter alip--1: Active-learning instruction in physics},}\ }\href@noop {}
  {\bibfield  {journal} {\bibinfo  {journal} {Am. J. Phys.}\ }\textbf {\bibinfo
  {volume} {80}},\ \bibinfo {pages} {478} (\bibinfo {year} {2012})}\BibitemShut
  {NoStop}%
\bibitem [{\citenamefont {Chasteen}\ \emph {et~al.}(2015)\citenamefont
  {Chasteen}, \citenamefont {Wilcox}, \citenamefont {Caballero}, \citenamefont
  {Perkins}, \citenamefont {Pollock},\ and\ \citenamefont
  {Wieman}}]{chasteen2015sei}%
  \BibitemOpen
  \bibfield  {author} {\bibinfo {author} {\bibfnamefont {Stephanie~V.}\
  \bibnamefont {Chasteen}}, \bibinfo {author} {\bibfnamefont {Bethany}\
  \bibnamefont {Wilcox}}, \bibinfo {author} {\bibfnamefont {Marcos~D.}\
  \bibnamefont {Caballero}}, \bibinfo {author} {\bibfnamefont {Katherine~K.}\
  \bibnamefont {Perkins}}, \bibinfo {author} {\bibfnamefont {Steven~J.}\
  \bibnamefont {Pollock}}, \ and\ \bibinfo {author} {\bibfnamefont {Carl~E.}\
  \bibnamefont {Wieman}},\ }\bibfield  {title} {\enquote {\bibinfo {title}
  {Educational transformation in upper-division physics: The science education
  initiative model, outcomes, and lessons learned},}\ }\href {\doibase
  10.1103/PhysRevSTPER.11.020110} {\bibfield  {journal} {\bibinfo  {journal}
  {Phys. Rev. ST Phys. Educ. Res.}\ }\textbf {\bibinfo {volume} {11}},\
  \bibinfo {pages} {020110} (\bibinfo {year} {2015})}\BibitemShut {NoStop}%
\bibitem [{\citenamefont {McDermott}\ and\ \citenamefont
  {Shaffer}(2001)}]{mcdermott2001tutorials}%
  \BibitemOpen
  \bibfield  {author} {\bibinfo {author} {\bibfnamefont {Lillian~C}\
  \bibnamefont {McDermott}}\ and\ \bibinfo {author} {\bibfnamefont {Peter~S}\
  \bibnamefont {Shaffer}},\ }\href@noop {} {\emph {\bibinfo {title} {Tutorials
  in introductory physics and homework package}}}\ (\bibinfo  {publisher}
  {Prentice Hall},\ \bibinfo {year} {2001})\BibitemShut {NoStop}%
\bibitem [{\citenamefont {Finkelstein}\ and\ \citenamefont
  {Pollock}(2005)}]{finkelstein2005tutorial}%
  \BibitemOpen
  \bibfield  {author} {\bibinfo {author} {\bibfnamefont {N.~D.}\ \bibnamefont
  {Finkelstein}}\ and\ \bibinfo {author} {\bibfnamefont {S.~J.}\ \bibnamefont
  {Pollock}},\ }\bibfield  {title} {\enquote {\bibinfo {title} {Replicating and
  understanding successful innovations: Implementing tutorials in introductory
  physics},}\ }\href {\doibase 10.1103/PhysRevSTPER.1.010101} {\bibfield
  {journal} {\bibinfo  {journal} {Phys. Rev. ST Phys. Educ. Res.}\ }\textbf
  {\bibinfo {volume} {1}},\ \bibinfo {pages} {010101} (\bibinfo {year}
  {2005})}\BibitemShut {NoStop}%
\bibitem [{\citenamefont {Caballero}\ \emph {et~al.}(2015)\citenamefont
  {Caballero}, \citenamefont {Wilcox}, \citenamefont {Doughty},\ and\
  \citenamefont {Pollock}}]{caballero2015mathphys}%
  \BibitemOpen
  \bibfield  {author} {\bibinfo {author} {\bibfnamefont {Marcos~D}\
  \bibnamefont {Caballero}}, \bibinfo {author} {\bibfnamefont {Bethany~R}\
  \bibnamefont {Wilcox}}, \bibinfo {author} {\bibfnamefont {Leanne}\
  \bibnamefont {Doughty}}, \ and\ \bibinfo {author} {\bibfnamefont {Steven~J}\
  \bibnamefont {Pollock}},\ }\bibfield  {title} {\enquote {\bibinfo {title}
  {Unpacking students’ use of mathematics in upper-division physics: where do
  we go from here?}}\ }\href@noop {} {\bibfield  {journal} {\bibinfo  {journal}
  {European Journal of Physics}\ }\textbf {\bibinfo {volume} {36}},\ \bibinfo
  {pages} {065004} (\bibinfo {year} {2015})}\BibitemShut {NoStop}%
\bibitem [{\citenamefont {Wilcox}\ \emph {et~al.}(2013)\citenamefont {Wilcox},
  \citenamefont {Caballero}, \citenamefont {Rehn},\ and\ \citenamefont
  {Pollock}}]{wilcox2013acer}%
  \BibitemOpen
  \bibfield  {author} {\bibinfo {author} {\bibfnamefont {Bethany~R.}\
  \bibnamefont {Wilcox}}, \bibinfo {author} {\bibfnamefont {Marcos~D.}\
  \bibnamefont {Caballero}}, \bibinfo {author} {\bibfnamefont {Daniel~A.}\
  \bibnamefont {Rehn}}, \ and\ \bibinfo {author} {\bibfnamefont {Steven~J.}\
  \bibnamefont {Pollock}},\ }\bibfield  {title} {\enquote {\bibinfo {title}
  {Analytic framework for students’ use of mathematics in upper-division
  physics},}\ }\href {\doibase 10.1103/PhysRevSTPER.9.020119} {\bibfield
  {journal} {\bibinfo  {journal} {Phys. Rev. ST Phys. Educ. Res.}\ }\textbf
  {\bibinfo {volume} {9}},\ \bibinfo {pages} {020119} (\bibinfo {year}
  {2013})}\BibitemShut {NoStop}%
\bibitem [{\citenamefont {Catrambone}(2011)}]{catrambone2011taps}%
  \BibitemOpen
  \bibfield  {author} {\bibinfo {author} {\bibfnamefont {Richard}\ \bibnamefont
  {Catrambone}},\ }\bibfield  {title} {\enquote {\bibinfo {title} {Task
  analysis by problem solving (taps): Uncovering expert knowledge to develop
  high-quality instructional materials and training},}\ }in\ \href@noop {}
  {\emph {\bibinfo {booktitle} {2011 Learning and Technology Symposium}}}\
  (\bibinfo {address} {Columbus, Georgia},\ \bibinfo {year} {2011})\BibitemShut
  {NoStop}%
\bibitem [{\citenamefont {Hammer}(2000)}]{hammer2000resources}%
  \BibitemOpen
  \bibfield  {author} {\bibinfo {author} {\bibfnamefont {David}\ \bibnamefont
  {Hammer}},\ }\bibfield  {title} {\enquote {\bibinfo {title} {Student
  resources for learning introductory physics},}\ }\href@noop {} {\bibfield
  {journal} {\bibinfo  {journal} {Am. J. Phys.}\ }\textbf {\bibinfo {volume}
  {68}},\ \bibinfo {pages} {S52--S59} (\bibinfo {year} {2000})}\BibitemShut
  {NoStop}%
\bibitem [{\citenamefont {Bing}(2008)}]{bing2008thesis}%
  \BibitemOpen
  \bibfield  {author} {\bibinfo {author} {\bibfnamefont {Thomas~J.}\
  \bibnamefont {Bing}},\ }\emph {\bibinfo {title} {An Epistemic Framing
  Analysis of Upper-Level Physics Students' Use of Mathematics}},\ \href@noop
  {} {\bibinfo {type} {Dissertation}},\ \bibinfo  {school} {University of
  Maryland} (\bibinfo {year} {2008})\BibitemShut {NoStop}%
\bibitem [{\citenamefont {Nguyen}\ and\ \citenamefont
  {Rebello}(2011)}]{nguyen2011int}%
  \BibitemOpen
  \bibfield  {author} {\bibinfo {author} {\bibfnamefont {Dong-Hai}\
  \bibnamefont {Nguyen}}\ and\ \bibinfo {author} {\bibfnamefont {N.~Sanjay}\
  \bibnamefont {Rebello}},\ }\bibfield  {title} {\enquote {\bibinfo {title}
  {Students' difficulties with integration in electricity},}\ }\href {\doibase
  10.1103/PhysRevSTPER.7.010113} {\bibfield  {journal} {\bibinfo  {journal}
  {Phys. Rev. ST Phys. Educ. Res.}\ }\textbf {\bibinfo {volume} {7}},\ \bibinfo
  {pages} {010113} (\bibinfo {year} {2011})}\BibitemShut {NoStop}%
\bibitem [{\citenamefont {Meredith}\ and\ \citenamefont
  {Marrongelle}(2008)}]{meredith2008resources}%
  \BibitemOpen
  \bibfield  {author} {\bibinfo {author} {\bibfnamefont {Dawn~C.}\ \bibnamefont
  {Meredith}}\ and\ \bibinfo {author} {\bibfnamefont {Karen~A.}\ \bibnamefont
  {Marrongelle}},\ }\bibfield  {title} {\enquote {\bibinfo {title} {How
  students use mathematical resources in an electrostatics context},}\ }\href
  {\doibase 10.1119/1.2839558} {\bibfield  {journal} {\bibinfo  {journal}
  {American Journal of Physics}\ }\textbf {\bibinfo {volume} {76}},\ \bibinfo
  {pages} {570--578} (\bibinfo {year} {2008})}\BibitemShut {NoStop}%
\bibitem [{\citenamefont {Hu}\ and\ \citenamefont
  {Rebello}(2013)}]{hu2013differentials}%
  \BibitemOpen
  \bibfield  {author} {\bibinfo {author} {\bibfnamefont {Dehui}\ \bibnamefont
  {Hu}}\ and\ \bibinfo {author} {\bibfnamefont {N.~Sanjay}\ \bibnamefont
  {Rebello}},\ }\bibfield  {title} {\enquote {\bibinfo {title} {Understanding
  student use of differentials in physics integration problems},}\ }\href
  {\doibase 10.1103/PhysRevSTPER.9.020108} {\bibfield  {journal} {\bibinfo
  {journal} {Phys. Rev. ST Phys. Educ. Res.}\ }\textbf {\bibinfo {volume}
  {9}},\ \bibinfo {pages} {020108} (\bibinfo {year} {2013})}\BibitemShut
  {NoStop}%
\bibitem [{\citenamefont {Amos}\ and\ \citenamefont
  {Heckler}(2018)}]{amos2018differentials}%
  \BibitemOpen
  \bibfield  {author} {\bibinfo {author} {\bibfnamefont {Nathaniel}\
  \bibnamefont {Amos}}\ and\ \bibinfo {author} {\bibfnamefont {Andrew~F.}\
  \bibnamefont {Heckler}},\ }\bibfield  {title} {\enquote {\bibinfo {title}
  {Mediating relationship of differential products in understanding integration
  in introductory physics},}\ }\href {\doibase
  10.1103/PhysRevPhysEducRes.14.010105} {\bibfield  {journal} {\bibinfo
  {journal} {Phys. Rev. Phys. Educ. Res.}\ }\textbf {\bibinfo {volume} {14}},\
  \bibinfo {pages} {010105} (\bibinfo {year} {2018})}\BibitemShut {NoStop}%
\bibitem [{\citenamefont {Savelsbergh}\ \emph {et~al.}(2011)\citenamefont
  {Savelsbergh}, \citenamefont {de~Jong},\ and\ \citenamefont
  {Ferguson-Hessler}}]{savelsbergh2011approach}%
  \BibitemOpen
  \bibfield  {author} {\bibinfo {author} {\bibfnamefont {Elwin~R.}\
  \bibnamefont {Savelsbergh}}, \bibinfo {author} {\bibfnamefont {Ton}\
  \bibnamefont {de~Jong}}, \ and\ \bibinfo {author} {\bibfnamefont {Monica
  G.~M.}\ \bibnamefont {Ferguson-Hessler}},\ }\bibfield  {title} {\enquote
  {\bibinfo {title} {Choosing the right solution approach: The crucial role of
  situational knowledge in electricity and magnetism},}\ }\href {\doibase
  10.1103/PhysRevSTPER.7.010103} {\bibfield  {journal} {\bibinfo  {journal}
  {Phys. Rev. ST Phys. Educ. Res.}\ }\textbf {\bibinfo {volume} {7}},\ \bibinfo
  {pages} {010103} (\bibinfo {year} {2011})}\BibitemShut {NoStop}%
\bibitem [{\citenamefont {Taylor}(2005)}]{taylor2005classmech}%
  \BibitemOpen
  \bibfield  {author} {\bibinfo {author} {\bibfnamefont {John~R.}\ \bibnamefont
  {Taylor}},\ }\href {http://books.google.com/books?id=P1kCtNr-pJsC} {\emph
  {\bibinfo {title} {Classical mechanics}}}\ (\bibinfo  {publisher} {University
  Science Books},\ \bibinfo {year} {2005})\BibitemShut {NoStop}%
\bibitem [{\citenamefont {Boas}(2006)}]{boas2006math}%
  \BibitemOpen
  \bibfield  {author} {\bibinfo {author} {\bibfnamefont {M.L.}\ \bibnamefont
  {Boas}},\ }\href {http://books.google.com/books?id=C3-NQgAACAAJ} {\emph
  {\bibinfo {title} {Mathematical Methods in the Physical Sciences}}}\
  (\bibinfo  {publisher} {John Wiley \& Sons},\ \bibinfo {year}
  {2006})\BibitemShut {NoStop}%
\bibitem [{\citenamefont {Griffiths}(1999)}]{griffiths1999em}%
  \BibitemOpen
  \bibfield  {author} {\bibinfo {author} {\bibfnamefont {David~J.}\
  \bibnamefont {Griffiths}},\ }\href
  {http://books.google.com/books?id=M8XvAAAAMAAJ} {\emph {\bibinfo {title}
  {Introduction to electrodynamics}}}\ (\bibinfo  {publisher} {Prentice Hall},\
  \bibinfo {year} {1999})\BibitemShut {NoStop}%
\bibitem [{\citenamefont {Mazur}(1997)}]{mazur1997pi}%
  \BibitemOpen
  \bibfield  {author} {\bibinfo {author} {\bibfnamefont {Eric}\ \bibnamefont
  {Mazur}},\ }\href@noop {} {\emph {\bibinfo {title} {Peer Instruction: A
  User's Manual}}},\ Series in Educational Innovation\ (\bibinfo  {publisher}
  {Prentice Hall},\ \bibinfo {address} {Upper Saddle River},\ \bibinfo {year}
  {1997})\BibitemShut {NoStop}%
\bibitem [{\citenamefont {Chasteen}\ \emph {et~al.}(2012)\citenamefont
  {Chasteen}, \citenamefont {Pollock}, \citenamefont {Pepper},\ and\
  \citenamefont {Perkins}}]{chasteen2012transforming}%
  \BibitemOpen
  \bibfield  {author} {\bibinfo {author} {\bibfnamefont {Stephanie~V}\
  \bibnamefont {Chasteen}}, \bibinfo {author} {\bibfnamefont {Steven~J}\
  \bibnamefont {Pollock}}, \bibinfo {author} {\bibfnamefont {Rachel~E}\
  \bibnamefont {Pepper}}, \ and\ \bibinfo {author} {\bibfnamefont
  {Katherine~K}\ \bibnamefont {Perkins}},\ }\bibfield  {title} {\enquote
  {\bibinfo {title} {Transforming the junior level: Outcomes from instruction
  and research in e\&m},}\ }\href@noop {} {\bibfield  {journal} {\bibinfo
  {journal} {Phys. Rev. ST Phys. Educ. Res.}\ }\textbf {\bibinfo {volume}
  {8}},\ \bibinfo {pages} {020107} (\bibinfo {year} {2012})}\BibitemShut
  {NoStop}%
\end{thebibliography}%

\end{document}